\documentclass[twocolumn,aps,prb,floats,superscriptaddress,showpacs]{revtex4}
\usepackage[latin1]{inputenc}
\usepackage{color}
\usepackage{textcomp}
\usepackage{amsmath}
\usepackage{amssymb}
\usepackage{graphicx}
\usepackage{esint}
\usepackage[unicode=true,pdfusetitle,
 bookmarks=true,bookmarksnumbered=false,bookmarksopen=false,
 breaklinks=false,pdfborder={0 0 1},backref=false,colorlinks=true]
 {hyperref}

\makeatletter
\@ifundefined{textcolor}{}
{%
 \definecolor{BLACK}{gray}{0}
 \definecolor{WHITE}{gray}{1}
 \definecolor{RED}{rgb}{1,0,0}
 \definecolor{GREEN}{rgb}{0,1,0}
 \definecolor{BLUE}{rgb}{0,0,1}
 \definecolor{CYAN}{cmyk}{1,0,0,0}
 \definecolor{MAGENTA}{cmyk}{0,1,0,0}
 \definecolor{YELLOW}{cmyk}{0,0,1,0}
}

\usepackage{epsfig}\usepackage{dcolumn}\usepackage{bm}\usepackage{amsfonts}

\makeatother

\begin{document}

\title{Mott-Anderson transition in disordered charge transfer model: \linebreak{}
insights from typical medium theory}

\author{W. S. Oliveira}

\affiliation{Departamento de F\'{i}sica, Universidade Federal de Minas Gerais, Avenida
Ant\^onio Carlos, 6627, Belo Horizonte, MG, Brazil}

\affiliation{Departamento de F\'{i}sica, Universidade Federal do Piau\'{i}, Campus 
Universit\'ario Ministro Petr\^onio Portella, Teresina, PI, Brazil}

\author{M. C. O. Aguiar%
\footnote{Correspondence should be addressed to aguiar@fisica.ufmg.br.%
}}

\affiliation{Departamento de F\'{i}sica, Universidade Federal de Minas Gerais, Avenida
Ant\^onio Carlos, 6627, Belo Horizonte, MG, Brazil}

\author{V. Dobrosavljevi\'{c}}

\affiliation{Department of Physics and National High Magnetic Field Laboratory,
Florida State University, Tallahassee, FL 32306, USA}

\begin{abstract}
The Mott-Anderson transition in the disordered charge-transfer model
displays several new features in comparison to what is found in the
disordered single-band Hubbard model, as recently demonstrated by
large-scale computational ({\it statistical} dynamical mean field
theory) studies. Here we show that a much simpler 
typical medium theory approach (TMT-DMFT) to the same model is able 
to capture most qualitative and even quantitative aspects of the phase 
diagram, the emergence of an intermediate electronic Griffiths phase, 
and the critical behavior close to the metal-insulator transition. 
Conceptual and mathematical simplicity of the TMT-DMFT formulation thus 
makes it possible to gain useful new insight into the mechanism of the 
Mott-Anderson transition in these models. 
\end{abstract}
\maketitle

\section{Introduction}

The physical mechanism behind the metal-insulator transition (MIT) remains
one of the basic science questions that still lack complete understanding,
both on the conceptual and the technical levels. Early work on the
subject focused on examining stability of the metallic phase with
respect to weak disorder,\cite{fink-jetp84,fink-zphys84,cclm-prb84}
within the framework of a quasiparticle picture and an appropriate
generalization\cite{ckl} of Landau's Fermi liquid theory. These
approaches, while formally elegant and appealing, suffer from several
conceptual shortcomings that render them of limited relevance to many
real materials. Essentially, these treatments describe situations
where disorder is viewed as the driving force for the metal-insulator
transition, and interactions only modify the details of the critical
behavior. What is implicitly \emph{assumed} within this picture is
that the ``host'' Fermi liquid is far from any interaction-induced
instabilities, where strong correlation effects may destroy\cite{RoP2005review}
the very existence of well-defined quasiparticles. Unfortunately,
recent experiments on several model systems, such as two-dimensional 
electron systems\cite{elihu_rmp} and doped semiconductors,\cite{mit_sib} 
have provided evidence that these strong correlation effects may very
well be the dominant driving force for electron localization, and
thus should be explicitly included in the theory. 



Both disorder (Anderson) and correlation (Mott) mechanisms to localization 
can be treated on the same foot by extensions of dynamical mean field 
theory\cite{dmftrev} (DMFT). In 
the so-called {\it statistical} DMFT\cite{statdmft} ({\it stat}DMFT) 
strong correlations are considered in a self-consistent DMFT fashion, while 
disorder fluctuations are treated by a (numerically) exact computational 
scheme. Because it is numerically very demanding, this method has been 
utilized only in a handful of theoretical studies of the Mott-Anderson 
transition.\cite{statdmft,eric,eric2,hofstetter-2010prb,hofstetter-2011prb} 
In particular, 
two of us have recently used it to study the precise form of quantum 
criticality of the charge transfer model.\cite{ctstat} A much simpler 
approach - the combination between typical-medium theory\cite{tmt} (TMT)
and DMFT - provided the first self-consistent description of the 
Mott-Anderson transition, and offered some insight into its critical 
regime.\cite{vollhardt,ourtmt} When applied to the Hubbard model, for 
weak to moderate disorder TMT-DMFT found a transition closely 
resembling the clean Mott point, while only at stronger disorder Anderson 
localization modified the critical behavior.\cite{vollhardt,ourtmt} 
Here, we employ TMT-DMFT to solve the charge transfer model, obtaining 
results in surprisingly good agreement with those recently obtained by 
us\cite{ctstat} within the more shophisticated {\it stat}DMFT method.

Besides describing the MIT, in this work we also address the 
electronic Griffiths phase with non-Fermi liquid behavior, which is 
experimentally observed in heavy-fermion systems,\cite{stewart1,stewart2} 
as well as in doped semiconductors.\cite{doped-book} 
In these systems, the susceptibility is seen to diverge in a power-law fashion
in the low temperature limit, not only in the insulating phase, but
also in the metallic side of the MIT.\cite{nfl_sip_b} In a number of 
systems, it is the disorder that is responsible for this non-Fermi
liquid behavior.\cite{RoP2005review} Theoretically, this phase is ``naturally'' 
incorporated in the description given by {\it stat}DMFT.\cite{localization03}
Within the DMFT 
framework, it can be addressed by considering the effective model 
proposed in Ref.~\onlinecite{effmodel}. By combining the latter with TMT, 
we are able to confirm that for the CT model a Griffiths phase is observed 
in the region just preceding the 
correlation induced MIT, as within {\it stat}DMFT.\cite{ctstat} 

In the present work, we consider the charge transfer (CT) model because 
it can describe the systems of our interest better than the single-band 
Hubbard model. This is the case since the former gives a more realistic 
description of spatial charge re-distribution as the MIT is approached, 
which is important because local correlation effects strongly depend on 
orbital occupation. In this context, it is interesting to note that the 
CT model phase diagram seems to differ from 
that of the single-band Hubbard, even in qualitative aspects.\cite{ctstat} 
Moreover, when both DMFT and TMT-DMFT are applied to the Hubbard model, 
the cavity field does not fluctuate, meaning that important fluctuation 
effects associated with the Griffiths phase 
and the precise nature of quantum criticality are ignored. 
In contrast, when the same method is used to solve the CT model a degree of 
fluctuation is retained, according to the effective model cited 
above,\cite{effmodel} hence even the simplified theories capture effects 
such as the Griffiths phase. Another advantage of considering 
the CT model is that 
its standard formulation uses the $U=\infty$ constraint for the correlated
band, allowing a simpler large $N$ (slave boson) solution,\cite{coleman} 
which is not available for finite $U$ Hubbard models.

As mentioned above, to solve the disordered CT model in this paper we use 
the TMT-DMFT method, which allows a detailed description of the system 
close to the MIT because of its conceptual and mathematical simplicity. 
According to our 
current results, as the interaction induced transition is approached, a 
fraction of sites turn into local moment, but {\it not all} of them do it.
This is in contrast to the TMT-DMFT results for the Hubbard model,\cite{ourtmt} 
where {\it all} sites turn into local moments close to the Mott transition.  
The phase diagram for the CT model thus includes a {\it disordered} Mott 
insulating phase which is qualitatively different than the Mott insulator 
observed for the Hubbard model. The disorder induced MIT is also qualitatively 
different than for the Hubbard model. For the CT
model most of the sites Anderson localize, but {\it none} of them turn into 
local moment as disorder increases. In the case of the Hubbard model, 
we have a two-fluid picture, where a fraction 
of the sites go through Anderson localization, while {\it the rest} of them 
Mott localize.\cite{ourtmt}

The paper is organized as follows. In the next section we define the
model we consider and the method we use to solve it. Section III is
devoted to our numerical results: we present our phase diagram,
discuss the disorder (section III.A) and the interaction (section III.B)
induced transitions, with special emphasis to the behavior of the physical 
quantities that characterize the transitions, and finally present results
related to the Griffiths phase (section III.C). We end by 
summarizing our main conclusions.

\section{The model and its solution}

\subsection{Charge transfer model and TMT-DMFT equations}

The CT model is a two band model, where one band represents conduction
electrons and the other corresponds to localized or $f$-type electrons,
for which the electron-electron interactions are strong. 
It has been used to describe various systems, including 
oxides\cite{ctm_oxides} and doped semiconductors;\cite{doped-book} for
the latter, the disordered version of the model is the relevant one, which
is indeed the problem we address in this paper. 

The CT model description of
the Mott transition can be understood as follows: in the clean case, 
the insulating phase is approached as the $f$-electron energy decreases, 
which implies in a smaller number of conduction electrons per site; the 
transition itself takes place when the latter vanishes. A careful study 
on the regimes where this model can be used to describe the Mott 
transition in the clean case can be found in Ref.~\onlinecite{ctmarcelo},
for example.

In the disordered case, the CT model is given
by the disordered Anderson lattice model supplemented by the condition
that the average number of electrons on each site is equal to $1$.
The Hamiltonian for the Anderson lattice model is 
\begin{eqnarray}
H & = & \sum_{ij\sigma}\left[(\varepsilon_{j}-\mu)\delta_{ij}-t\right]c_{i\sigma}^{\dagger}c_{j\sigma}+(E_{f}-\mu)\sum_{j\sigma}f_{j\sigma}^{\dagger}f_{j\sigma}\nonumber \\
 & + & V\sum_{j\sigma}\left(c_{j\sigma}^{\dagger}f_{j\sigma}+f_{j\sigma}^{\dagger}c_{j\sigma}\right)+U\sum_{j}n_{fj\uparrow}n_{fj\downarrow},\label{hand}
\end{eqnarray}
where $c_{j\sigma}^{\dagger}$ ($c_{j\sigma}$) creates (destroys)
a conduction electron with spin $\sigma$ on site $j$, $f_{j\sigma}^{\dagger}$
and $f_{j\sigma}$ are the corresponding creation and annihilation
operators for a localized $f$-electron with spin $\sigma$ on site
$j$, $n_{fj\sigma}=f_{j\sigma}^{\dagger}f_{j\sigma}$ is the number
operator for $f$-electrons, $t$ is the hopping amplitude to nearest
neighbors, $E_{f}$ is the $f$-electron energy, $U$ is the on-site
repulsion between $f$-electrons, $V$ is the hybridization between
conduction and $f$-electrons, and $\mu$ is the chemical potential.
Throughout this paper we use the half-bandwidth for conduction electrons
as the unit of energy; the hybridization potential is chosen to be
$V=0.5$.

In eq.~(\ref{hand}), disorder is introduced through the on-site
energies $\varepsilon_{j}$ for conduction electrons, which follow
a distribution $P(\varepsilon)$. As we want to be able to address
the electronic Griffiths phase, we must deserve special attention
to the disorder distribution we consider. As we mentioned before,
this phase appears naturally when one treats the disordered correlated
system through \textit{stat}DMFT,\cite{localization03} but this is
not necessarily the case when standard DMFT is considered. In this
case, it has been shown that essentially all the properties of the
electronic Griffiths phase can be described \textit{if} we ``correctly''
choose the model to study and the disorder distribution.\cite{effmodel}

According to Ref.~\onlinecite{effmodel}, the recipe to describe
the Griffiths phase in a DMFT level is to include in the calculation
cavity fluctuations naturally described within \textit{stat}DMFT.
Firstly, one has to consider a two band model as the CT model we treat
here: in this case the bath seen by each impurity problem fluctuates,
that is, changes from site to site {[}see eq.~(\ref{delta_result})
below{]}. In addition, the disorder should be present in the on-site
conduction electron energy, which necessarily follows a Gaussian distribution.
This specific form of disorder generates a distribution of \textit{renormalized}
energies which is also Gaussian, as it is the case when \textit{stat}DMFT
with \textit{any} disorder distribution of \textit{bare} energies
is considered. Following these findings, in this paper we assume a
Gaussian form for $P(\varepsilon)$, with zero mean and standard deviation
equal to $W$, to be able to describe the Griffiths phase within TMT-DMFT.

To finish the description of the CT model, we add that the condition
that the average number of electrons on each site is equal to $1$
can be enforced by adjusting the chemical potential and can be written
as 
\begin{equation}
\langle n_{cj}\rangle+\langle n_{fj}\rangle=1,
\end{equation}
where $n_{fj}=n_{fj\uparrow}+n_{fj\downarrow}$ gives the number of
$f$-electrons on site $j$, $n_{cj}=n_{cj\uparrow}+n_{cj\downarrow}$
is the corresponding number operator for conduction electrons, with
$n_{cj\sigma}=c_{j\sigma}^{\dagger}c_{j\sigma}$, and the averages
are taken over the distribution $P(\varepsilon)$.

As anticipated in the Introduction, we use a combination of TMT and
DMFT to solve the disordered CT model. Within this combination,\cite{tmt,vollhardt,ourtmt}
the lattice problem is mapped onto an ensemble of single-impurity
problems, corresponding to sites with different values of the local
energy $\varepsilon_{j}$, each being embedded in a typical effective
medium which is self-consistently calculated. In contrast to standard
DMFT,\cite{screening} TMT-DMFT determines this effective medium by
replacing the spectrum of the environment (``cavity'') for each
site by its typical value, which is determined by the process of \textit{geometric}
averaging.

To be more specific, within TMT-DMFT the Hamiltonian of eq.~(\ref{hand})
is mapped onto an ensemble of single-impurity problems, each of which
is given by the following action 
\begin{eqnarray}
 &  & S(j)\label{acaoef}\\
 &  & =\sum_{\sigma}\int_{0}^{\beta}d\tau\int_{0}^{\beta}d\tau'f_{j\sigma}^{\dagger}(\tau)\left[\delta(\tau-\tau')\left(\partial_{\tau}+E_{f}-\mu\right)\right.\nonumber \\
 &  & +\left.\Delta_{fj}(\tau-\tau')\right]f_{j\sigma}(\tau')+U\int_{0}^{\beta}d\tau n_{fj\uparrow}(\tau)n_{fj\downarrow}(\tau),\nonumber 
\end{eqnarray}
where the Fourier transform of $\Delta_{fj}(\tau-\tau')$ satisfies
\begin{equation}
\Delta_{fj}(i\omega)=\frac{V^{2}}{i\omega+\mu-\varepsilon_{j}-t^{2}G_{c}^{typ}(i\omega)}.\label{delta_result}
\end{equation}
A Bethe lattice of infinite coordination number was considered when
writing the above equation.

$G_{c}^{typ}(i\omega)$ is the typical Green's function for conduction
electrons, which within TMT-DMFT is given by the Hilbert transform
of $\rho_{c}^{typ}(\omega)$, the typical value of the local density-of-states
(LDOS). In equations, we have 
\begin{equation}
G_{c}^{typ}(i\omega)=\int_{-\infty}^{\infty}d\omega^{\prime}\frac{\rho_{c}^{typ}(\omega^{\prime})}{i\omega-\omega^{\prime}},
\end{equation}
where 
\begin{equation}
\rho_{c}^{typ}(\omega)=\exp\{\left<\ln\rho_{cj}(\omega)\right>\}\label{rhotyp}
\end{equation}
and 
\begin{equation}
\rho_{cj}(\omega)=-\pi^{-1}\operatorname{Im}G_{cj}(\omega)\label{ldos}
\end{equation}
is the LDOS.

The local Green's function for conduction electrons appearing in the
above equation satisfies 
\begin{equation}
G_{cj}(i\omega)=\frac{1}{i\omega+\mu-\varepsilon_{j}-t^{2}G_{c}^{typ}(i\omega)-\Phi_{j}(i\omega)},\label{condauto}
\end{equation}
where 
\begin{equation}
\Phi_{j}(i\omega)=\frac{V^{2}}{i\omega+\mu-E_{f}-\Sigma_{fj}(i\omega)},\label{phi}
\end{equation}
and $\Sigma_{fj}(i\omega)$ is the single-impurity self-energy, which
is a solution of the action given in eq.~(\ref{acaoef}).

By looking at eq.~(\ref{rhotyp}), for example, one can conclude
that the problem defined by these equations corresponds to a self-consistent
calculation. In other words, within TMT-DMFT the conduction electron
effective medium seen by each impurity is self-consistently determined.

\subsection{Slave-boson impurity solver}

To solve the single-impurity problems of eq. (\ref{acaoef}), we use
the slave-boson (SB) technique in the $U\rightarrow\infty$ limit.\cite{readnewns,coleman}
In this case, the impurity Green's function can be written as 
\begin{eqnarray}
G_{fj}(i\omega) & = & \frac{Z_{j}}{i\omega-\varepsilon_{fj}-Z_{j}\Delta_{fj}(i\omega)}\\
 & \equiv & Z_{j}G_{fj}^{QP}(i\omega),
\end{eqnarray}
where $Z_{j}$ is the local quasiparticle (QP) weight and $\varepsilon_{fj}$
is the renormalized $f$-electron energy. These two parameters are
obtained by solving the following set of equations 
\begin{equation}
2\int_{0}^{\infty}\frac{d\omega}{\pi}\mbox{Re}\left[\Delta_{fj}(i\omega)G_{fj}^{QP}(i\omega)\right]=E_{f}-\varepsilon_{fj},
\end{equation}
\begin{equation}
Z_{j}+2\int_{0}^{\infty}\frac{d\omega}{\pi}\mbox{Re}\left[G_{fj}^{QP}(i\omega)\right]=0.
\end{equation}
For more details on the $U\rightarrow\infty$ SB treatment we refer
the reader to Ref.~\onlinecite{kdm}.

Before finishing the section, it is convenient to note that in terms
of the two SB parameters eq.~(\ref{phi}) can be rewritten as 
\begin{equation}
\Phi_{j}(i\omega)=\frac{Z_{j}V^{2}}{i\omega-\varepsilon_{fj}}.\label{phi2}
\end{equation}

\begin{figure}
\begin{centering}
\includegraphics[scale=0.33]{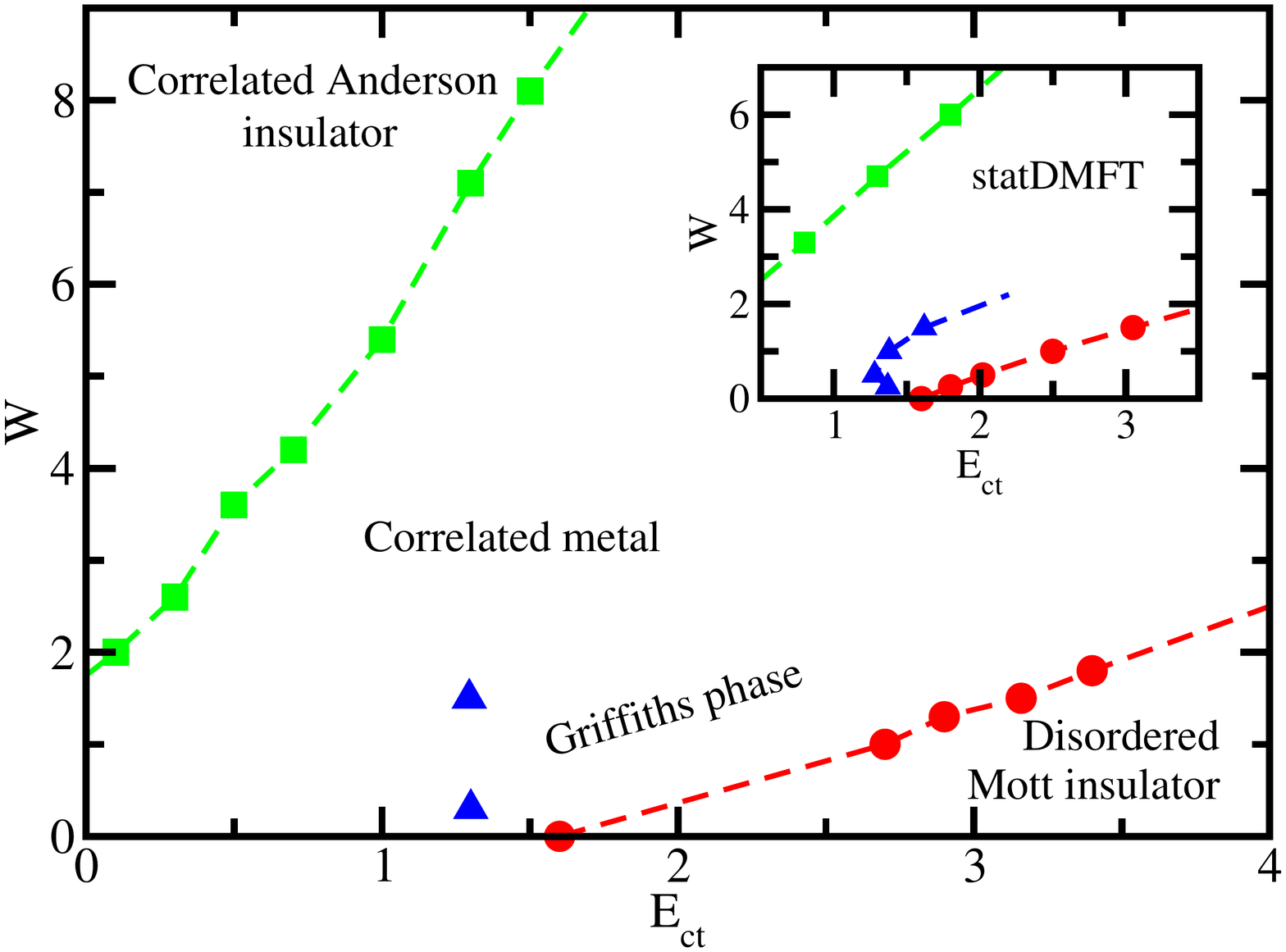} 
\caption{(Color online) Phase diagram of the disordered CT model obtained within
TMT-DMFT. $E_{ct}=-E_{f}$ is the CT energy and plays the role of
the Hubbard $U$. For comparison, the inset reproduces the results
obtained within \textit{stat}DMFT and presented in Ref.~\onlinecite{ctstat}.}
\label{fig1}
\par\end{centering}
\end{figure}

\section{Numerical results}

Let us now present and discuss the numerical results we obtained for
the CT model using TMT-DMFT. In this section, we also compare these
results with those obtained by two of us within the more sophisticated
\textit{stat}DMFT,\cite{ctstat} which provides an exact numerical
treatment of localization in the absence of interactions, and reduces
to the standard DMFT treatment in the absence of disorder.\cite{statdmft}

Fig.~\ref{fig1} presents our phase diagram. As we described previously,\cite{ctstat}
starting from a disordered correlated metal, a transition to a correlated
Anderson insulator takes place as disorder increases; on the other
hand, a disordered Mott insulating phase is observed for large values
of the CT energy. The latter is defined as $E_{ct}=-E_{f}$ and plays
the role of the interaction energy $U$ in the Hubbard model. By comparing
the results in the main panel of Fig.~\ref{fig1} with those in the
inset, we can see that in the case of the Mott-like transition TMT-DMFT
predicts the phase boundary in very good agreement with \textit{stat}DMFT.
For the Anderson transition, according to TMT-DMFT a slightly larger
amount of disorder than that observed in \textit{stat}DMFT is necessary
to drive the transition.

In the following, we look at how the order parameter and other physical
quantities behave as the transitions are approached.

\begin{figure}
\begin{centering}
\includegraphics[scale=0.33]{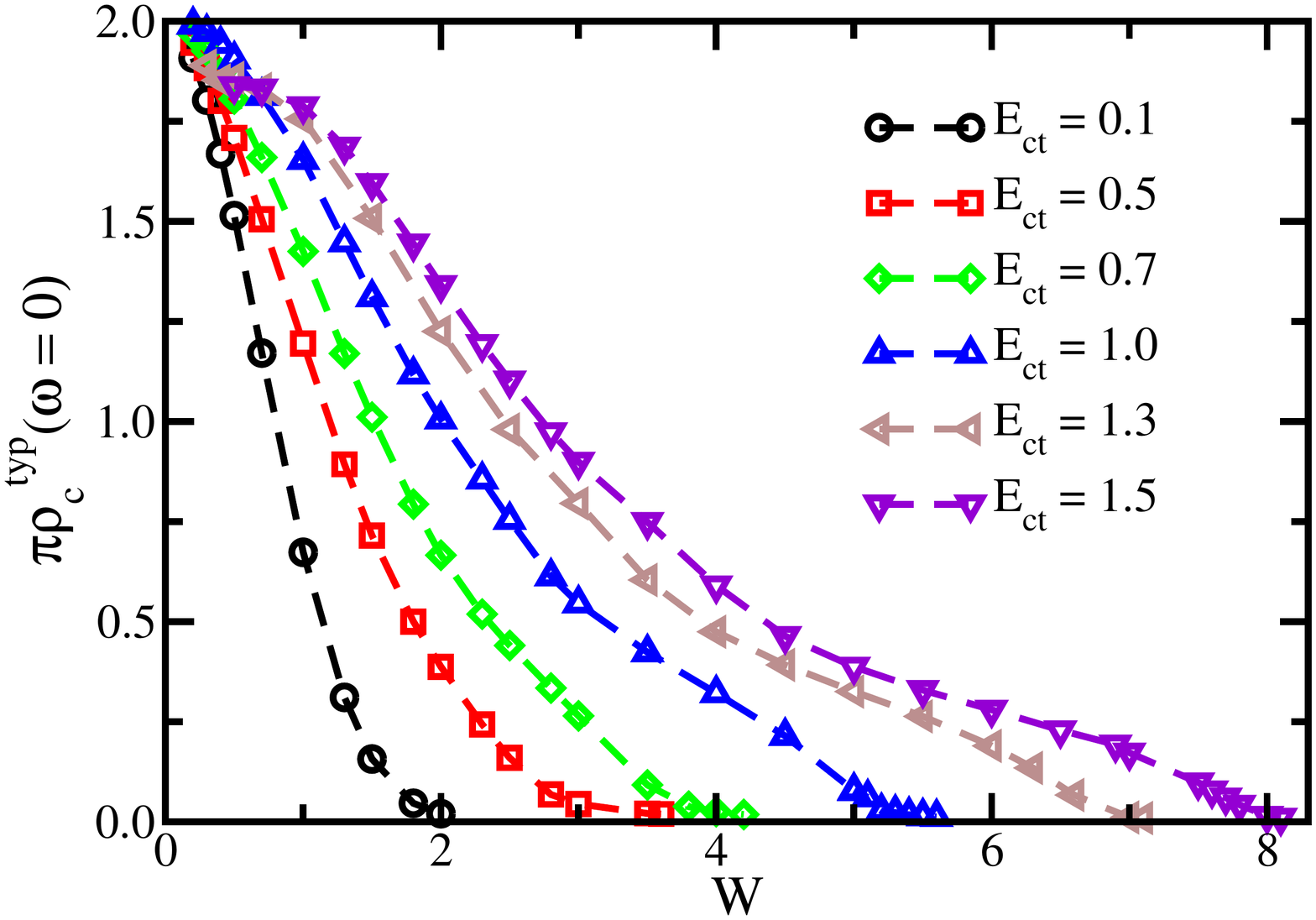} 
\caption{(Color online) Typical values of the LDOS for conduction electrons
at the Fermi energy as a function of the disorder strength, $W$,
for different values of the CT energy, $E_{ct}$, obtained within
TMT-DMFT.}
\label{fig2} 
\par\end{centering}
\end{figure}

\subsection{Disorder-driven transition}

\subsubsection{Critical behavior of the local density-of-states}

Fig.~\ref{fig2} shows the typical LDOS for conduction electrons
at the Fermi energy as the disorder-driven transition is approached
for different values of the CT energy. As expected, $\rho_{c}^{typ}(\omega=0)$
decreases from the clean value as $W$ increases, due to disorder
induced localization effects. The typical LDOS for conduction electrons
corresponds indeed to an order parameter within TMT-DMFT: its vanishing
defines the critical disorder at which the MIT takes place. In the
present case, for all values of $E_{ct}$, $\rho_{c}^{typ}(\omega=0)$
is seen to go to zero continuously as the MIT is approached, in agreement
with \textit{stat}DMFT results.\cite{ctstat}

\begin{figure}
\begin{centering}
\includegraphics[scale=0.33]{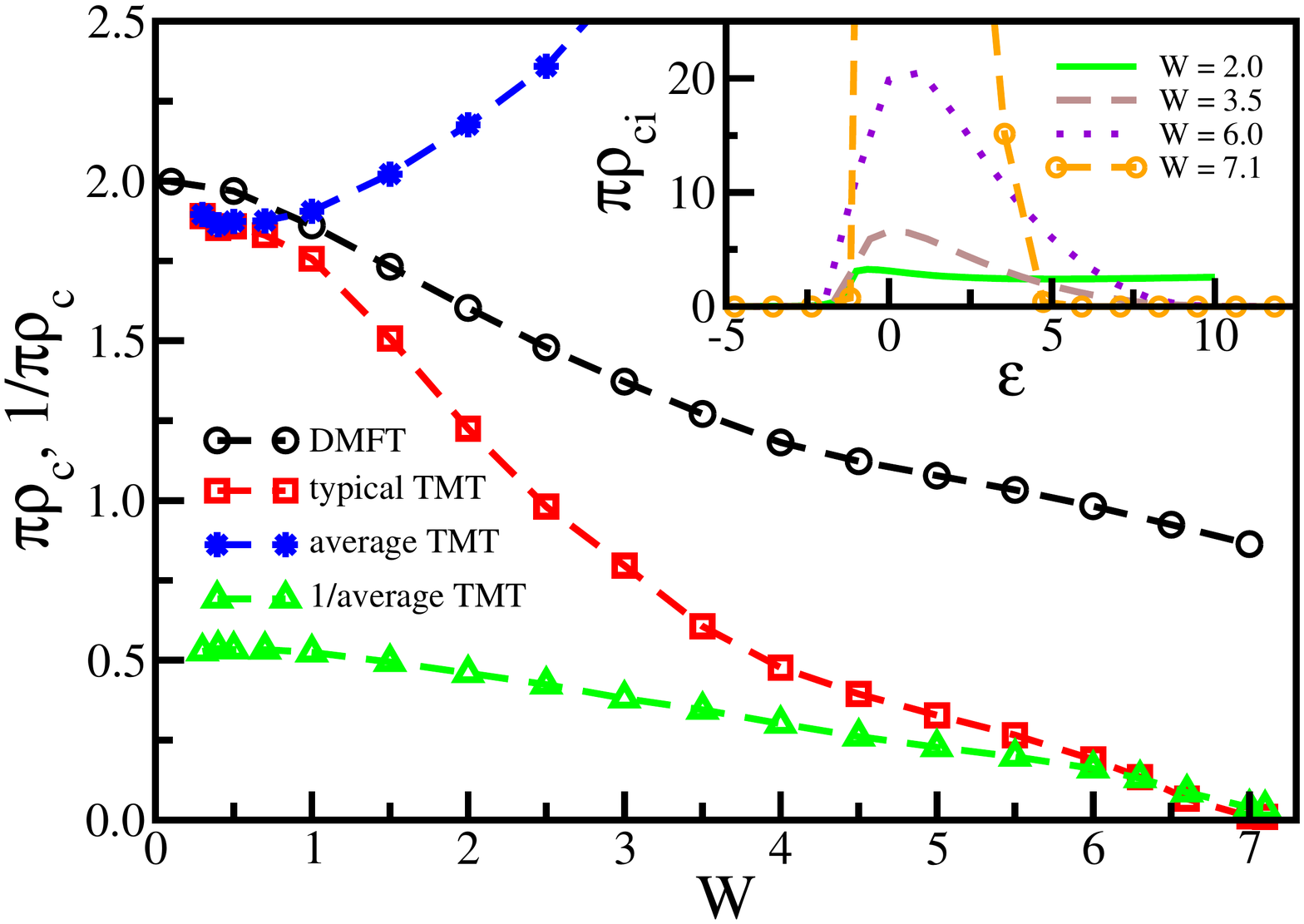} 
\caption{(Color online) Comparison between TMT-DMFT and \emph{stat}DMFT: results
are shown for the typical and 1/average values of the LDOS for conduction
electrons at the Fermi energy obtained for $E_{ct}=1.3$ and different
values of the disorder strength, $W$. The inset shows the typical
values of the QP weight $Z$ as a function of $W$ for both treatments.}
\label{fig3}
\par\end{centering}
\end{figure}

A detailed comparison with \textit{stat}DMFT results for $E_{ct}=1.3$
can be seen in Fig.~\ref{fig3}. In accordance with the phase diagram
of Fig.~\ref{fig1}, within TMT-DMFT the transition is seen at a
larger $W$ value than within \textit{stat}DMFT. Although in the present
treatment the bath fluctuates from site-to-site - note that the bath
given by eq.~(\ref{delta_result}) does depend on the site $j$,
our results in Fig.~\ref{fig3} may suggest that not all the fluctuations
induced by Anderson localization effects are captured by the simple
TMT-DMFT treatment. Still, the behavior of the typical (and the inverse
of average) LDOS is very similar in both treatments, allowing us to
conclude that TMT-DMFT does give a reasonable picture of the transition.
In addition, as pointed out before, since the latter is numerical
and analytically simpler, it facilitates the understanding of the
physics behind the problem we are looking at, as we discuss in this
paper for the CT model.

\begin{figure}[t]
\begin{centering}
\includegraphics[scale=0.33]{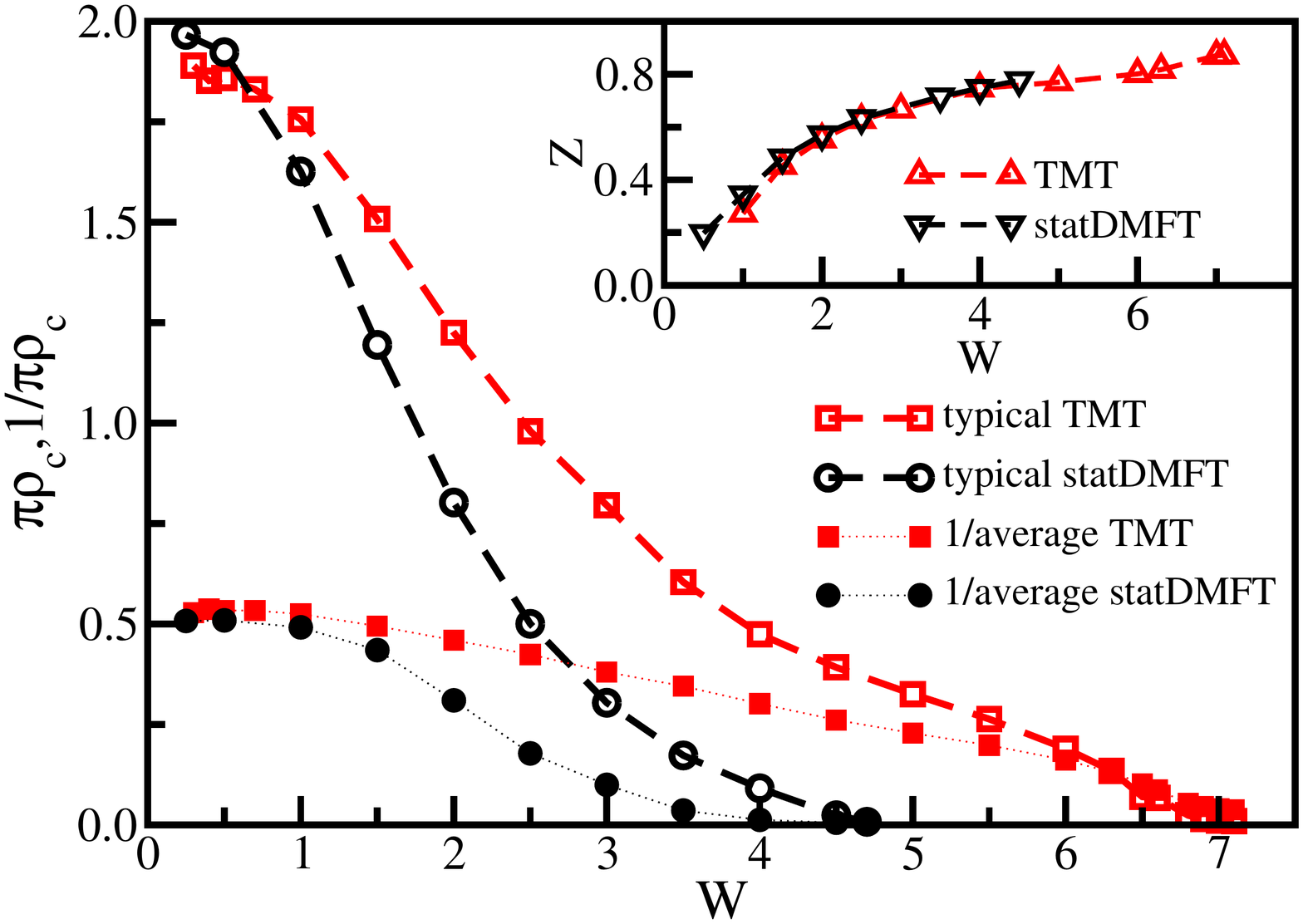} 
\caption{(Color online) Comparison between TMT-DMFT and CPA-DMFT: results are
shown for the typical and average values of the LDOS for conduction
electrons at the Fermi energy obtained for $E_{ct}=1.3$ and different
values of the disorder strength, $W$. The inset shows the TMT-DMFT
LDOS corresponding to each single-impurity problem as a function of
the on-site energy $\varepsilon$.}
\label{fig4} 
\par\end{centering}
\end{figure}

In Fig.~\ref{fig4} the behavior of the typical LDOS is compared
to that of the (arithmetic) average LDOS. It is interesting to note
that, as it is the general case for \textit{stat}DMFT results\cite{statdmft,ctstat}
(see also Fig.~\ref{fig3}), within TMT-DMFT the \textit{inverse}
of the average LDOS goes to zero at the same disorder at which $\rho_{c}^{typ}(\omega=0)$
vanishes. Fig.~\ref{fig4} also shows the results of standard CPA-DMFT
calculation, where disorder is treated as in the Coherent Potential
Approximation (CPA), being unable to describe Anderson localization
effects. 

\subsubsection{Statistics of local quasi-particle parameters}

Let us now look at the properties of the single-impurity problems
into which the lattice Hamiltonian is mapped. The inset of Fig.~\ref{fig4}
shows the LDOS for each single-impurity of the ensemble, which is
given by eq.~(\ref{ldos}) and from which the TMT-DMFT results in
the main figure are calculated. The conduction electrons, whose LDOS
we are analyzing, see the $f$-electrons through the function $\Phi(i\omega)$
{[}see eq.~(\ref{condauto}){]}, which we can identify as an effective
disorder potential. According to eq.~(\ref{phi2}), $\Phi(i\omega)$
is written in terms of the two Fermi liquid parameters, $Z$ and $\varepsilon_{f}$.
To understand the behavior of the LDOS close to the transition, below
we present and analyze the results for these two parameters.

\begin{figure}
\begin{centering}
\includegraphics[scale=0.33]{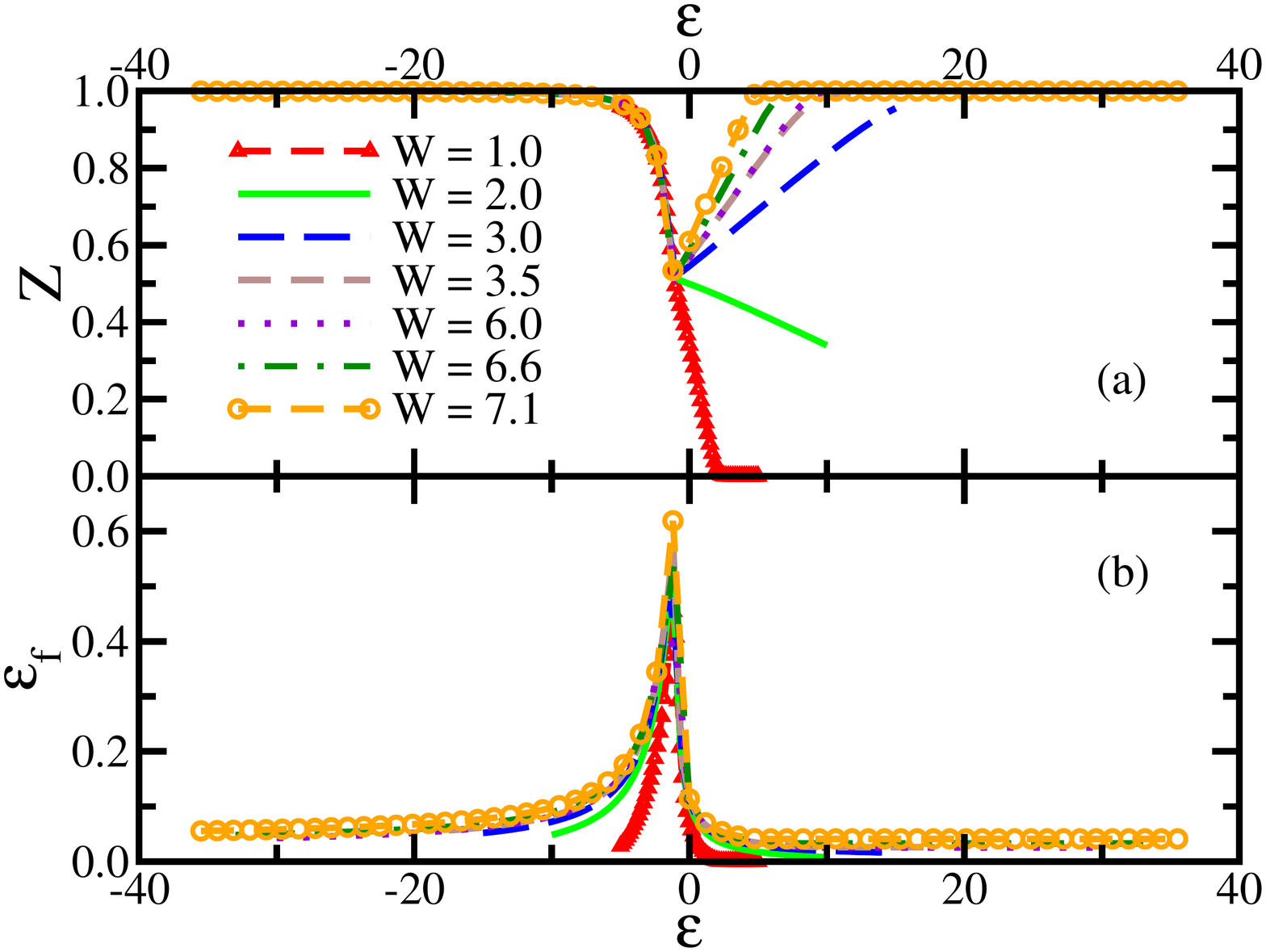} 
\caption{(Color online) (a) Quasiparticle weight $Z$ and (b)~renormalized
energy $\varepsilon_{f}$ as a function of the on-site energy $\varepsilon$
as the transition is approached (increase of $W$), for $E_{ct}=1.3$.
The results were obtained using TMT-DMFT.}
\label{fig5} 
\par\end{centering}
\end{figure}

Fig.~\ref{fig5}(a) shows the behavior of the QP weight $Z_{j}$
as a function of the on-site energy $\varepsilon_{j}$ as disorder
increases for $E_{ct}=1.3$ (same parameter of Fig.~\ref{fig3} and~\ref{fig4}).
For the smallest $W$ considered, the values of $Z_{j}$ are small
(there are even sites with $Z_{j}\rightarrow0$); as disorder increases,
the values of $Z_{j}$ increase, since the system tends to have most
of the sites with $Z_{j}=1$ (we know\cite{statdmft} that a ``pure''
Anderson insulator has all sites with $Z_{j}=1$). The site with $\varepsilon=E_{f}=-E_{ct}$
is a special one: it corresponds to $n_{c}=n_{f}=0.5$, which implies
in $Z=0.5$, as $Z_{j}=1-n_{fj}$ within SB. As a consequence of the
presence of the latter, close to the transition most of the sites
have $Z_{j}=1$, but some of them have $0.5\le Z_{j}<1$. Note that
none form local moments ($Z_{j}=0$), a situation completely different
than the one we will analyze in section III.B.2 below.

The typical values of $Z$ corresponding to the results in Fig.~\ref{fig5}(a)
are compared to the \textit{stat}DMFT results in the inset of Fig.~\ref{fig3}.
In the region where both treatments predict the system to be metallic,
although the typical LDOS is larger within TMT-DMFT than within \textit{stat}DMFT,
the typical values of $Z$ practically coincide.

The results for the second SB parameter - the renormalized energy
$\varepsilon_{fj}$ - are shown in Fig.~\ref{fig5}(b). The latter
is maximum for $\varepsilon_{j}=E_{f}=-E_{ct}$ and is relatively
small for the majority of the sites, which correspond indeed to the
intermediate to large $|\varepsilon_{j}|$ sites that have $Z_{j}=1$
close to the transition.

But which mechanism of localization dominates the current MIT? In
the case of the CT model, we have two kinds of electrons, localized
or $f$-electrons and conduction or $c$-electrons. Within the SB
method we consider, $Z_{j}=1-n_{fj}$ and, for the CT model, $\left<n_{cj}\right>+\left<n_{fj}\right>=1$.
According to the above, most of the sites have $Z_{j}=1$ (and small
$\varepsilon_{fj}$) close to the MIT. $n_{fj}=0$ for these sites,
corresponding to electrons occupying $c$-states, which are known
to localize as disorder increases. Indeed, $\rho_{cj}(\omega=0)\sim0$
for these sites, as one can see in the inset of Fig.~\ref{fig4}.
Thus, as the transition is approached, most of the sites go through
Anderson type of localization. In other words, in the present case
it is the Anderson mechanism for localization that is responsible
for driving the system as a whole through the MIT.

If we now compare the results in Fig.~\ref{fig5} for $W=3.5$ and
$W=6.0$, we see that $Z_{j}$ and $\varepsilon_{fj}$ coincide in
the range of $\varepsilon_{j}$ values present for both disorder strengths
(the range of $\varepsilon_{j}$ is, of course, larger for $W=6.0$
than for $W=3.5$), although $\rho_{cj}(\omega=0)$ do change in this
interval, as can be seen in the inset of Fig.~\ref{fig4}. Note,
however, that the rate at which the typical DOS decrease is smaller
in the region where $Z_{j}$ and $\varepsilon_{fj}$ coincide than
it is the case for smaller $W$ values. By looking at the different
quantities that determine $G_{cj}(\omega=0)$ {[}see eq.~(\ref{condauto}){]},
the results in Fig.~\ref{fig5} suggest that it is the bare disorder
($\varepsilon_{j}$) itself, rather than the scattering coming from
the $f$-electrons through $\Phi_{j}(\omega=0)$, that dominates the
behavior of the LDOS as the disorder driven-transition is approached
within TMT-DMFT.

To finish this subsection, the situation described here can be compared
to that observed within TMT-DMFT for the Hubbard model,\cite{ourtmt}
where close to the transition the sites have either $Z_{j}=1$ or
$Z_{j}=0$, corresponding to electrons going through Anderson or Mott
localization, respectively. (Note that the dependence of the LDOS
for conduction electrons on $Z$ is different in the two models considered.)
In the current case, we do not have sites going through Mott localization
($Z_{j}=0$): according to Fig.~\ref{fig5}(a), the majority of the
them have $Z_{j}=1$, while a finite fraction has $0.5\le Z_{j}<1$.
For the latter $n_{fj}\neq0$, that is, the occupation of \textit{strongly
correlated} $f$-electrons is different from zero for these sites.
Thus, although we do not have Mott localized sites, because of the
presence of the $0.5\le Z<1$ sites, the current situation is also
different than that in the non-interacting limit (where \textit{all}
sites have $Z_{j}=1$), and a \textit{correlated} Anderson insulator
is present in the CT model phase diagram.

After discussing the results for the disorder-induced transition,
in the next section we focus on the Mott-like transition.

\subsection{Interaction-driven transition}

\subsubsection{Critical behavior of the local density-of-states}

\begin{figure}
\begin{centering}
\includegraphics[scale=0.33]{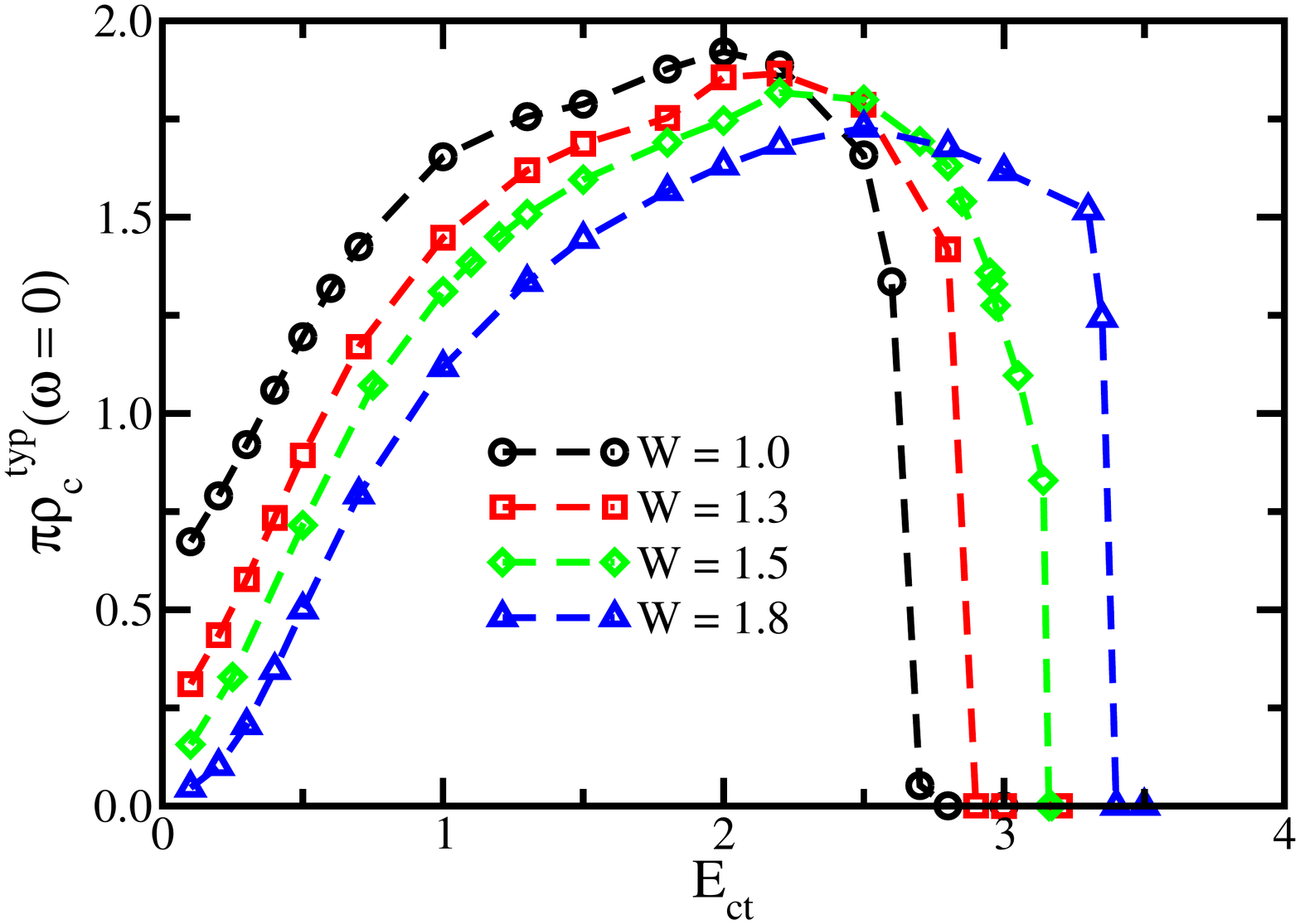} 
\caption{(Color online) Typical values of the LDOS for conduction electrons
at the Fermi energy as a function of the CT energy, $E_{ct}$, for
different values of the disorder strength, $W$, obtained within TMT-DMFT.}
\label{fig6} 
\par\end{centering}
\end{figure}

\begin{figure}
\begin{centering}
\includegraphics[scale=0.33]{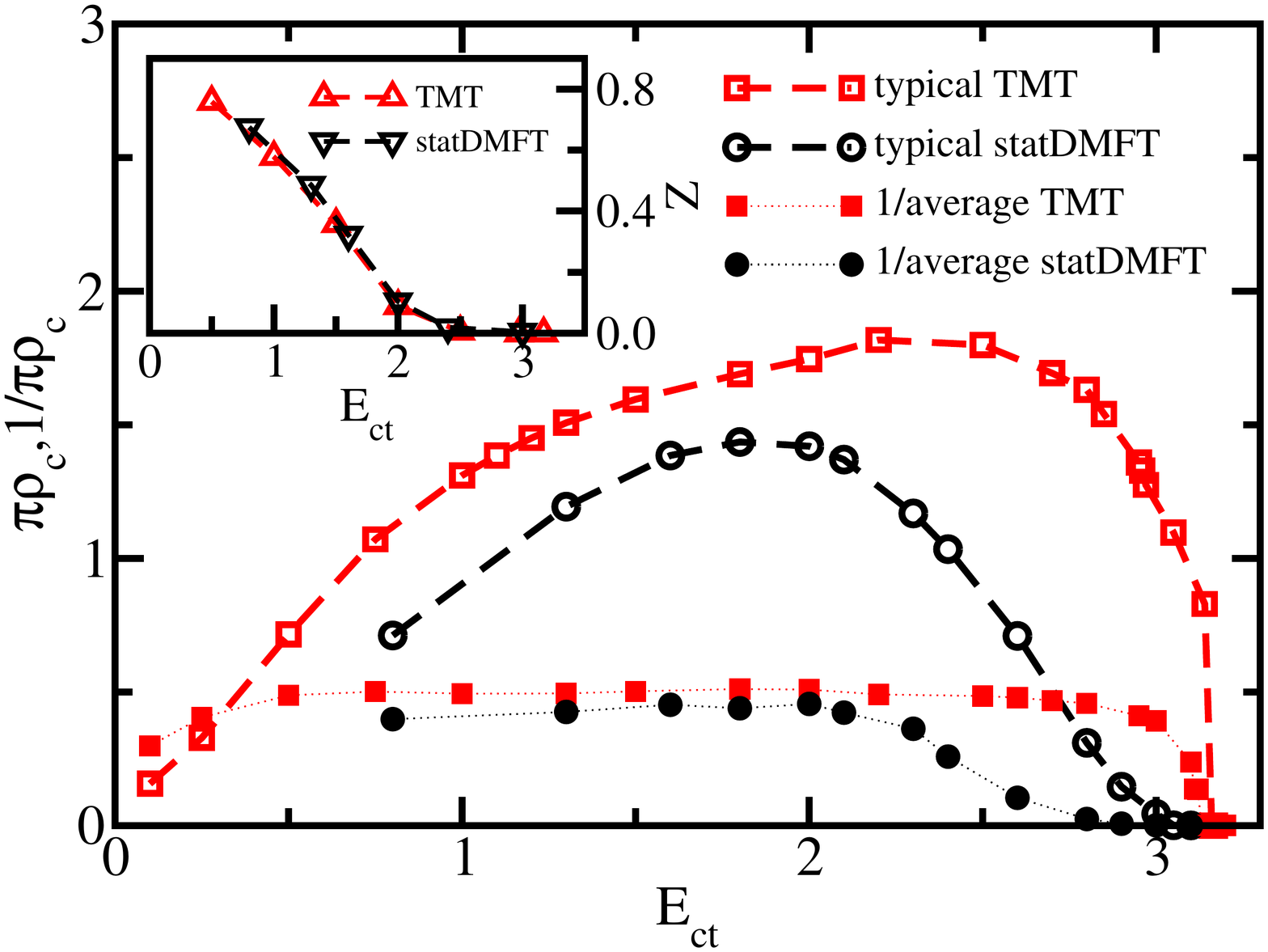} 
\caption{(Color online) Comparison between TMT-DMFT and \textit{stat}DMFT:
results are shown for the typical and 1/average values of the LDOS
for conduction electrons at the Fermi energy obtained for $W=1.5$
and different values of the CT energy, $E_{ct}$. The inset shows
the typical values of the QP weight $Z$ as a function of $E_{ct}$
for both treatments.}
\label{fig7}
\par\end{centering}
\end{figure}

Fig.~\ref{fig6} shows the typical LDOS for conduction electrons
at the Fermi energy as the Mott-like transition is approached for
intermediate values of disorder. A non-monotonic behavior is observed,
implying in an initial increase of the system ``conductivity'' when
$E_{ct}$ increases, which suggests that the disorder potential is
\textit{screened} by the correlation effects considered to exist between
$f$-electrons. This non-monotonic behavior is in agreement with the
\textit{stat}DMFT results we presented recently.\cite{ctstat} Indeed,
in the current case the screening is stronger than within \textit{stat}DMFT
- see, for example, the detailed comparison between TMT-DMFT and \textit{stat}DMFT
presented in Fig.~\ref{fig7} for $W=1.5$. Although strong, here
the screening is not perfect and $\rho_{typ}$ does not reach the
value corresponding to the clean limit, as it is the case for the
Hubbard model within DMFT\cite{screening} and TMT-DMFT.\cite{ourtmt}

Probably as a consequence of the strong disorder screening discussed
above, the typical DOS within TMT-DMFT is seen to present a jump at
the transition (see Fig.~\ref{fig6} and \ref{fig7}). This is in
disagreement with \textit{stat}DMFT for the CT model, which predicts
that the order parameter vanishes continuously as the transition is
approached.\cite{ctstat} Note, though, that according to Fig.~\ref{fig7}
a good agreement is observed between the two calculations concerning
the overall behavior of the typical and inverse of average LDOS, as
well as the $E_{ct}$ value at which the transition takes place (see
also Fig.~\ref{fig1}). Although the current results suggest that
TMT-DMFT does not completely describe Anderson localization effects,
which were shown to be responsible for the critical behavior also
in the vicinity of the Mott-like transition,\cite{ctstat} we can
say that it does give a reasonable picture of it.

\begin{figure}
\begin{centering}
\includegraphics[scale=0.33]{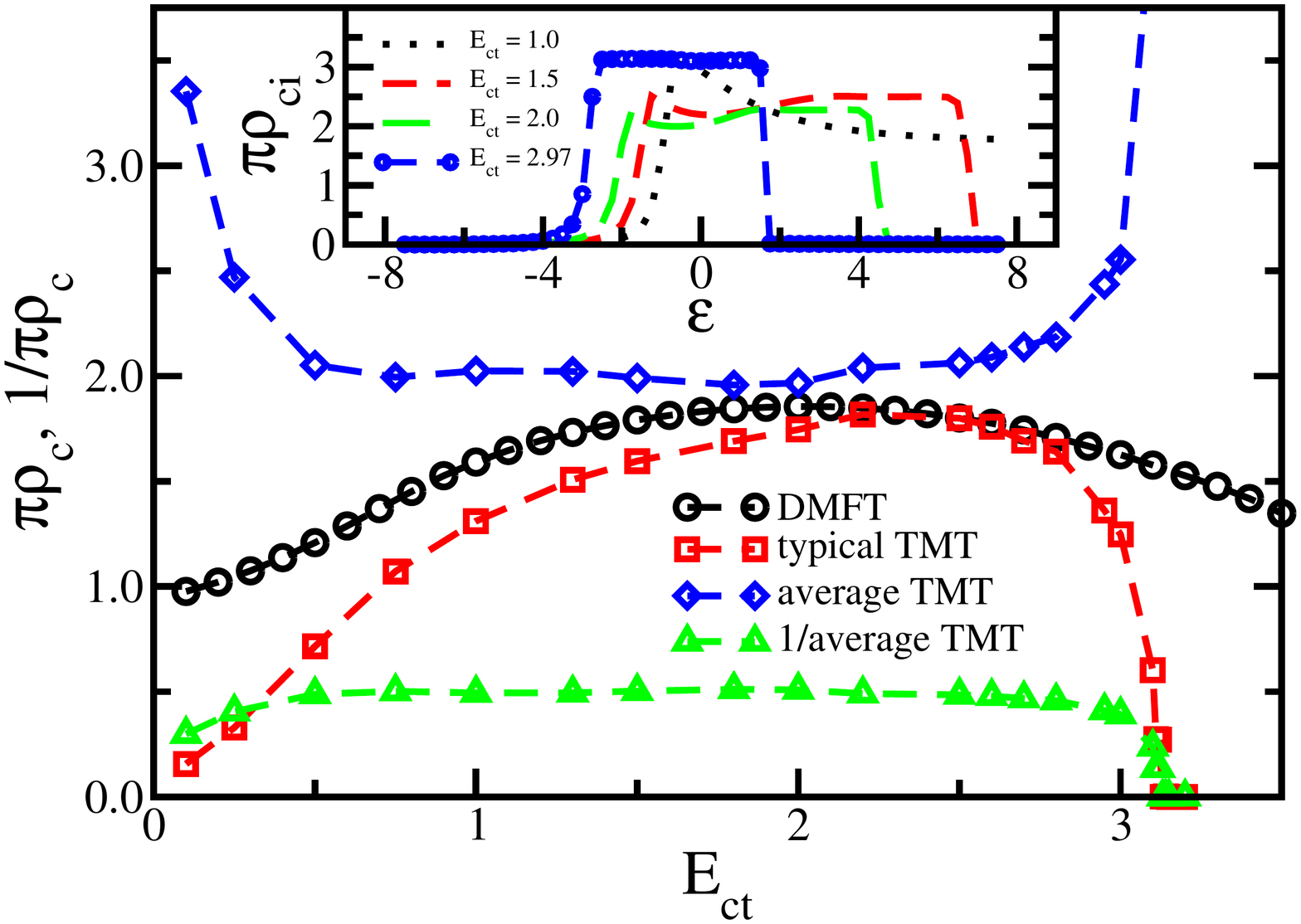} 
\caption{(Color online) Comparison between TMT-DMFT and standard DMFT: results
are shown for the typical and average values of the LDOS for conduction
electrons at the Fermi energy obtained for $W=1.5$ and different
values of the CT energy, $E_{ct}$. The inset shows the TMT-DMFT LDOS
as a function of the on-site energy $\varepsilon$ corresponding to
each single-impurity problem.}
\label{fig8} 
\par\end{centering}
\end{figure}

\begin{figure}
\begin{centering}
\includegraphics[scale=0.33]{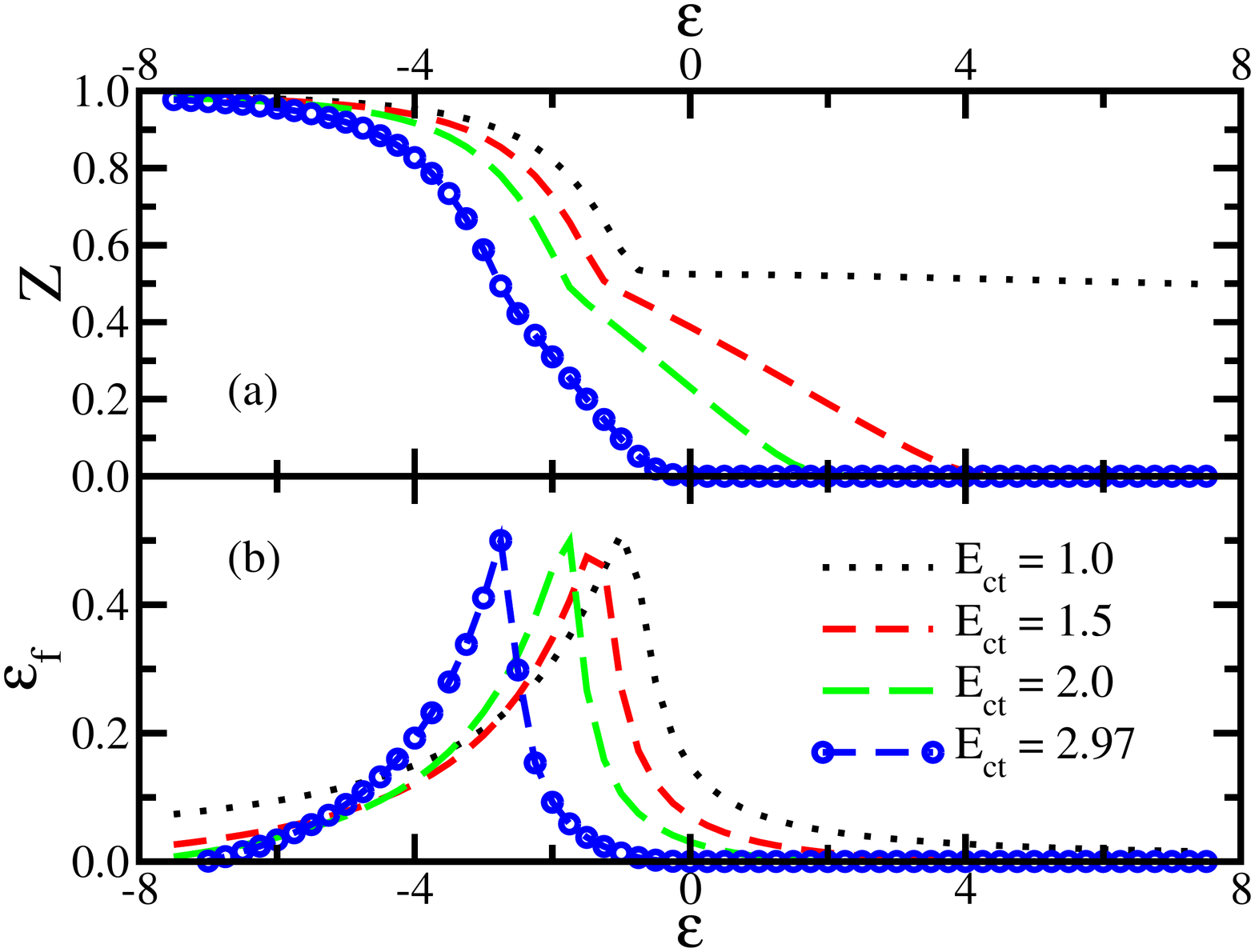} 
\caption{(Color online) (a) Quasiparticle weight $Z$ and (b) renormalized
energy $\varepsilon_{f}$ as a function of the on-site energy $\varepsilon$
as the transition is approached (increase of $E_{ct}$), for $W=1.5$.
Results were obtained using TMT-DMFT.}
\label{fig9}
\par\end{centering}
\end{figure}

To complete the discussion on the LDOS results, in Fig.~\ref{fig8}
we compare the typical and average values of the LDOS obtained within
TMT-DMFT with those valid within standard DMFT for $W=1.5$ (same
parameter as Fig.~\ref{fig7}). As it is the case for the disorder-induced
transition (see Fig.~\ref{fig4}), here the inverse of the average
LDOS within TMT-DMFT is seen to vanish together with the typical LDOS.
Also, standard DMFT average LDOS remains finite at the critical $E_{ct}$
predicted by TMT-DMFT.

\subsubsection{Statistics of local quasi-particle parameters}

As we did in the previous subsection, we now look at the properties
of the single-impurity problems, with the goal of understanding which
sites of the ensemble dominate the behavior of the LDOS in the critical
region. Fig.~\ref{fig9}(a) shows the QP weight $Z_{j}$ for each
single-impurity problem of the ensemble, for fixed disorder ($W=1.5$),
as the Mott-like transition is approached. As we can see, as $E_{ct}$
increases, the large $\varepsilon_{j}$ sites start to have $Z_{j}=0$;
as $E_{ct}$ increases even further, more sites present $Z_{j}=0$,
while the region of sites with $Z_{j}\neq0$ ($0<Z_{j}<1$ indeed)
shrinks to the left of the figure. Very close to the transition all
sites with positive $\varepsilon_{j}$, as well as few with $\varepsilon_{j}\precsim0$,
have $Z_{j}=0$. Correspondingly, the typical value of $Z$ decreases
as the transition is approached, in very good agreement with \textit{stat}DMFT,
as can be seen in the inset of Fig.~\ref{fig7}. Regarding the renormalized
energy, which is shown in Fig.~\ref{fig9}(b), we can see that the
sites that form local moments ($Z_{j}=0$) close to the transition
are completely screened ($\varepsilon_{fj}=0$). For the rest of the
sites, $\varepsilon_{fj}$ presents a non-monotonic behavior: it is
finite for intermediate, negative values of the bare energy and tends
to zero for the smallest $\varepsilon_{j}$ considered.

To understand the results described above, let us first analyze the
clean limit. In this case, DMFT maps the lattice problem onto only
one single-impurity problem - that with $\varepsilon=0$, which has
to satisfy $n_{c}+n_{f}=1$, within the CT model. The Mott transition
is approached as $E_{f}=-E_{ct}$ decreases, which favors the occupation
of the \textit{localized} $f$-level, implying in a decrease of the
occupation for conduction electrons, $n_{c}$; the transition happens
when $n_{c}=0$. Within the SB method, $Z=1-n_{f}$, which means that
$Z\rightarrow0$ as the transition is approached. As disorder is turned
on, an ensemble of single-impurity problems has to be solved; close
to the MIT transition, not all sites, but most of them, including
those around $\varepsilon_{j}=0$ (the one that remains in the clean
limit), have $Z_{j}\rightarrow0$, as can be seen in Fig.~\ref{fig9}(a).
These sites go through the Mott mechanism for localization; as they
are the majority in the present case, we conclude that Mott localization
dominates the MIT that happens as the CT energy increases.

The current situation is different than that observed for the Hubbard
model within TMT-DMFT.\cite{ourtmt} In the latter \textit{all} sites
turn to local moments as the transition is approached, in contrast
to the present case where there exist sites with $0<Z_{j}<1$. Because
of the presence of the $Z_{j}\neq0$ sites, the insulating phase we
observe here corresponds to a \textit{disordered} Mott insulator.

\begin{figure}[t]
\begin{centering}
\includegraphics[scale=0.33]{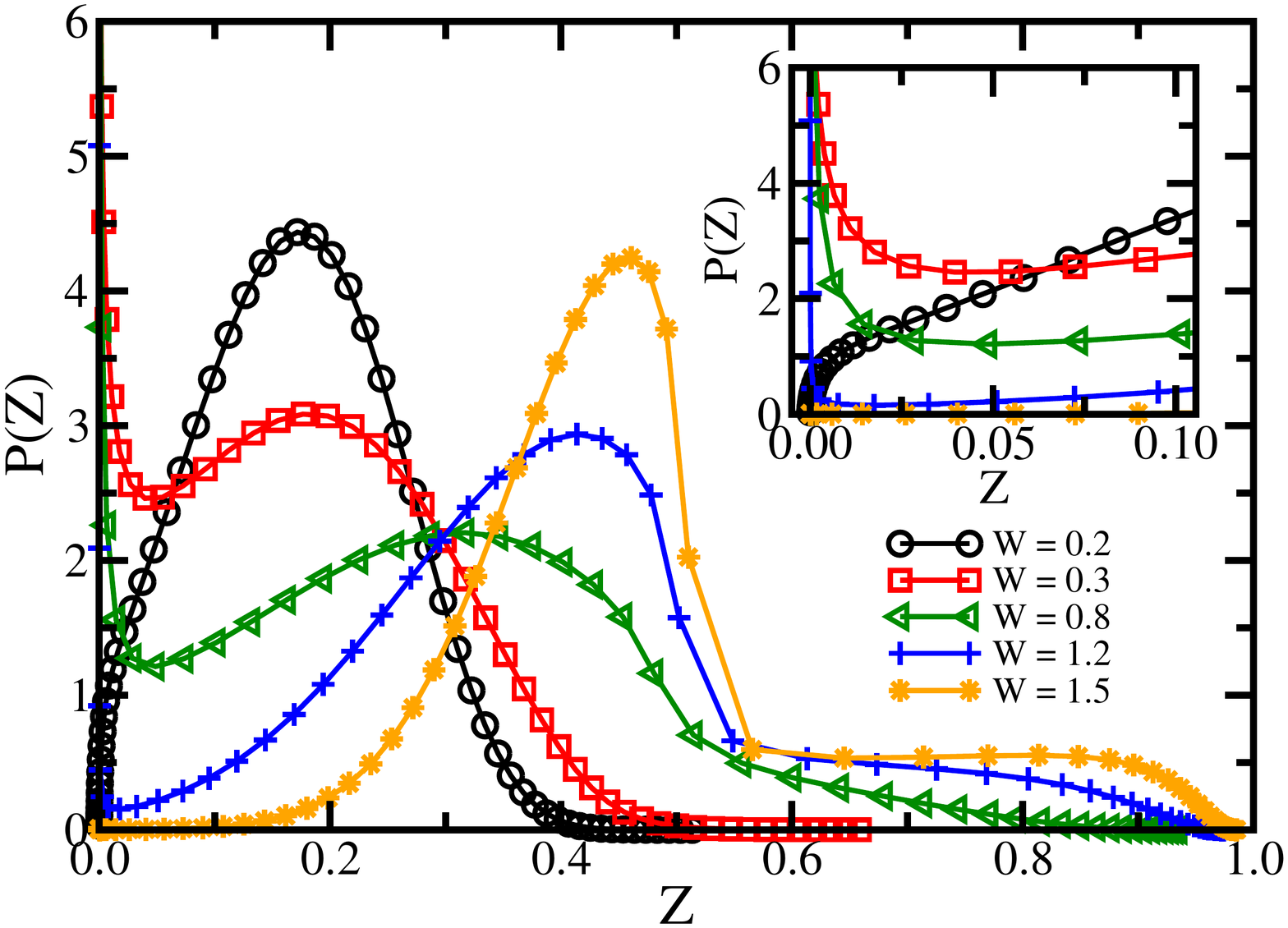} 
\caption{(Color online) Distribution of $Z$ obtained within TMT-DMFT as disorder
increases for $E_{ct}=1.3$. The inset highlights the fact that $P(Z=0)$
becomes different from zero for intermediate values of disorder, which
gives rise to a Griffiths phase in this range of parameters.}
\label{fig10} 
\par\end{centering}

\end{figure}

\subsection{Griffiths phase}

Besides giving a good description of the MIT, TMT-DMFT is also able
to describe the emergence of a Griffiths phase inside the disordered
metallic region. The latter is possible by considering a Gaussian
distribution of the on-site energy, as suggested in Ref.~\onlinecite{effmodel}
and summarized by us in Section II.

\begin{figure}[t]
\begin{centering}
\includegraphics[scale=0.33]{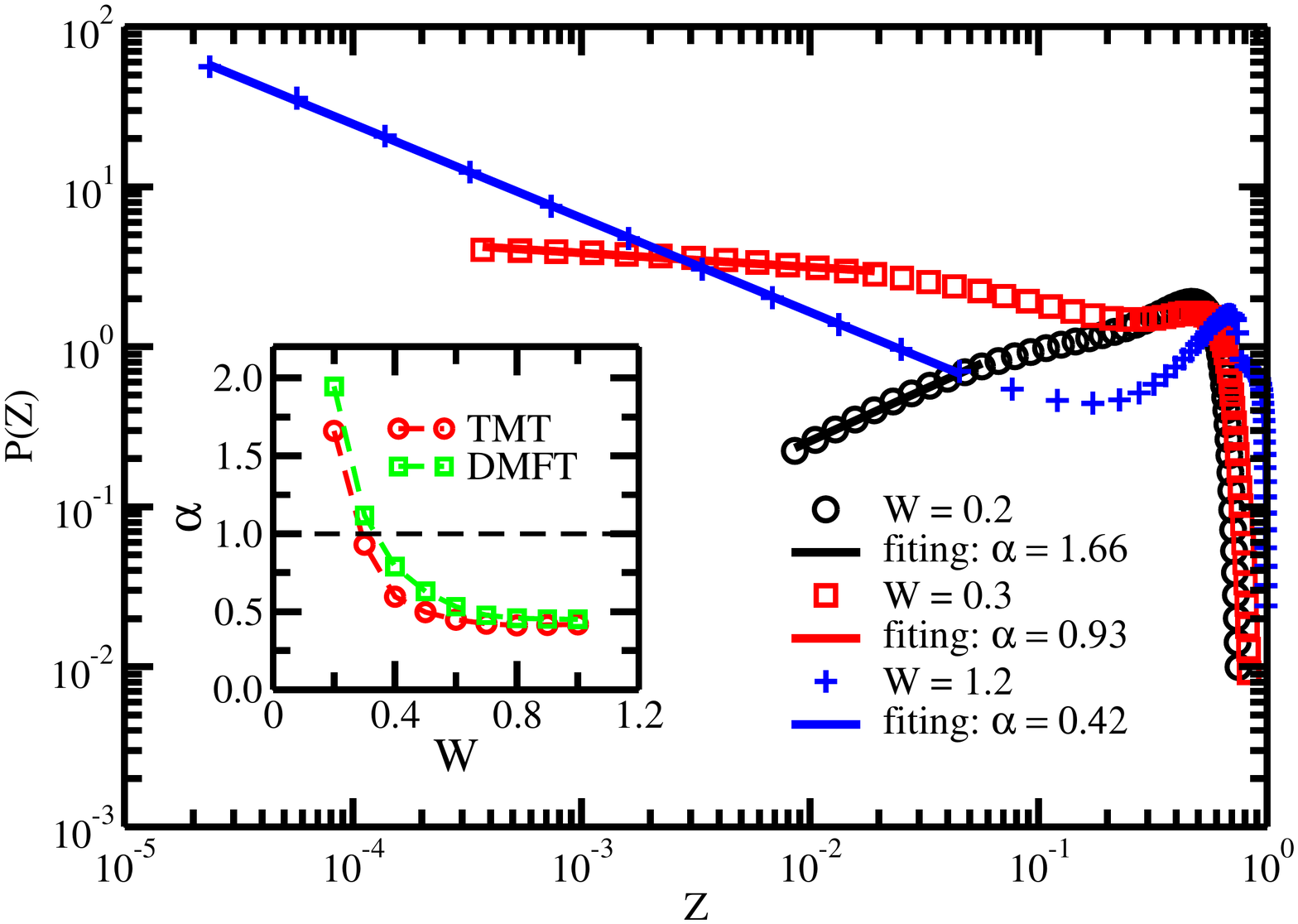} 
\caption{(Color online) Distribution of $Z$ and respective fits to a power-law
observed within TMT-DMFT for $E_{ct}=1.3$ and three values of $W$.
The inset presents the power-law exponent $\alpha$ as a function
of disorder, both within TMT-DMFT (as those in the main panel) and
standard DMFT (not shown).}
\label{fig11} 
\par\end{centering}
\end{figure}

To study the Griffiths phase, we focus on the behavior of $Z$ for
small disorder. In addition, instead of looking at its behavior as
a function of $\varepsilon$, as we did above, we look at the evolution
of its distribution, $P(Z)$. In Fig.~\ref{fig10} we have the results
for fixed CT energy, $E_{ct}=1.3$. As disorder increases, the distribution
moves to smaller values of $Z$. More importantly, it develops a tail
that follows a power-law of the form 
\begin{equation}
P(Z)\sim Z^{\alpha-1},\label{pl}
\end{equation}
which is better visualized in Fig.~\ref{fig11}. The exponent $\alpha$
is found by fitting the numerical data to the above equation; the
values obtained in the present case of $E_{ct}=1.3$ are shown in
the inset of the figure. As we can see, $\alpha$ is a continuous
function of $W$, becoming smaller than $1$ for $W\sim0.3$ in the
current case. As a consequence of $P(Z)$ following a power-law with
$\alpha<1$ (in some range of $W$), the system susceptibility and
specific heat divided by the temperature $T$ diverge in the low $T$
limit (see Ref.~\onlinecite{localization03} for a detailed discussion
on this). This characterizes a Griffiths phase with non-Fermi liquid
behavior.

According to the results in Fig.~\ref{fig10}, as disorder increases
even further, $P(Z)$ moves to larger values of $Z$ and the low $Z$
tail disappears. To precisely determine at which disorder the Griffiths
phase terminates for $E_{ct}=1.3$, one has to explore it in more
detail, for example by performing the current analysis as a function
of $E_{ct}$ for different, fixed $W$. This is illustrated below
for one fixed value of $W$.

\begin{figure}[t]
\begin{centering}
\includegraphics[scale=0.33]{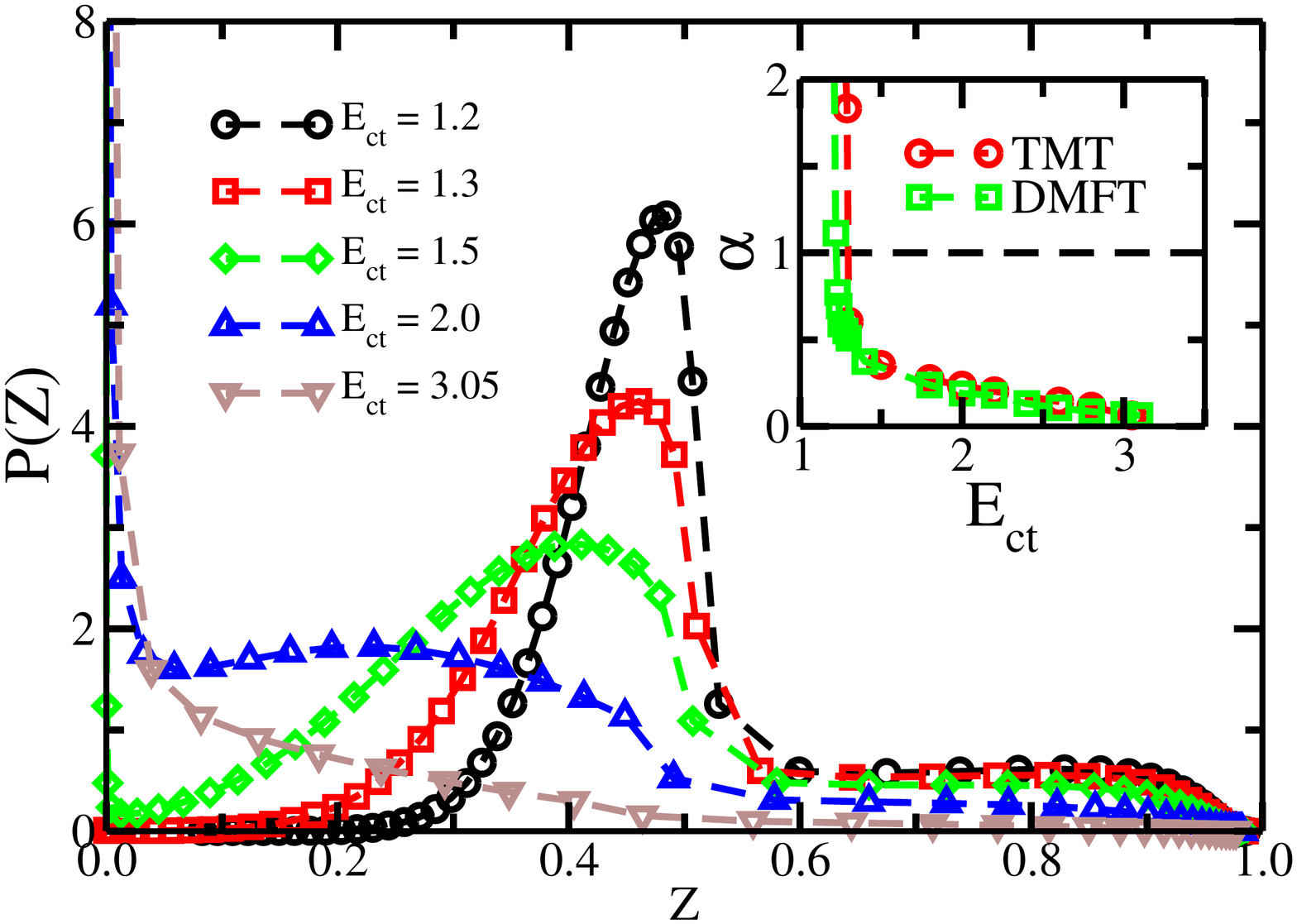} 
\caption{(Color online) Distribution of $Z$ obtained within TMT-DMFT as the
CT energy increases, for $W=1.5$. The inset presents the power-law
exponent $\alpha$ as a function of the CT energy obtained within
TMT-DMFT (main panel) and standard DMFT (not shown).}
\label{fig12} 
\par\end{centering}
\end{figure}

Fig.~\ref{fig12} shows the distribution of $Z$ (main panel) and
corresponding $\alpha$ (inset of the figure) for $W=1.5$. As $E_{ct}$
increases, $P(Z)$ moves to smaller values of $Z$. In the present
case, $\alpha$ becomes smaller than $1$ and the system enters the
Griffiths phase for $E_{ct}\sim1.3$. Differently than the previous
case, here $P(Z)$ moves to even smaller values of $Z$, with $\alpha$
decreasing to zero, as the Mott-like transition is approached. Note
that the $E_{ct}$ we have just found for the onset of the Griffiths
phase for $W=1.5$ corresponds to the $E_{ct}$ analyzed in Fig.~\ref{fig10}
and Fig.~\ref{fig11}; we can thus conclude that for $E_{ct}=1.3$
the Griffiths phase is observed between $W\sim0.3$ and $W\sim1.5$.

The results in these three figures indicate that within TMT-DMFT the
range of $W$ and $E_{ct}$ for which $\alpha<1$ corresponds to the
existence of a Griffiths phase in the region just preceding the Mott
transition. This region is signalized in the phase diagram of Fig.~\ref{fig1}
and is in accordance with \textit{stat}DMFT results for the same model
(see Ref.~\onlinecite{ctstat} and also the inset of Fig.~\ref{fig1}).
A similar behavior has also been observed within \textit{stat}DMFT
for the two-dimensional Hubbard model.\cite{eric}

To finish, in the insets of Fig.~\ref{fig11} and Fig.~\ref{fig12},
we compare the results obtained for $\alpha$ using TMT-DMFT (corresponding
to $P(Z)$ in the respective main panels) and standard DMFT. A Gaussian
distribution of disorder is used in the two calculations. In both
figures, in the range of parameters shown, a very good agreement is
seen between the two treatments considered. Note, however, that standard
DMFT agrees well with TMT-DMFT concerning the \textit{onset} of the
Griffiths phase, but not its \textit{extension}, as the former does
not give a good prediction for the critical $E_{ct}$ and $W$ values
at which the transitions take place, as previously showed in this
paper.

\section{Conclusions}

In this paper we solved the disordered charge-transfer model (CT)
by using an extension of dynamical mean field theory able to describe
Anderson localization effects. In general, our results compare surprisingly
well with those previously obtained by two of us using the \textit{stat}DMFT
treatment.\cite{ctstat} The current calculation is simpler than the
latter, allowing us to better characterize the system when the metal-insulator
transition is approached. Our results show, in particular, that as
the interaction induced transition is approached, a fraction of sites
turn into local moment, but \textit{not all} of them do it; this means
that the corresponding insulating phase is a \textit{disordered} Mott
insulator. In the case of the transition due to disorder, most of
the sites Anderson localize; some of the correlated sites, though,
remains occupied, corresponding to the presence of a \textit{correlated}
Anderson insulator in the phase diagram of the CT model.

In addition, according to our current TMT-DMFT results, the inverse
of the arithmetic local DOS is seen to vanish precisely at the disorder
or interaction value at which the typical local DOS goes to zero,
which indeed determines where the transition takes place. Exactly
the same behavior is observed within \textit{stat}DMFT,\cite{ctstat,statdmft}
but an explanation for it is yet not known. The fact that the current
treatment, which is analytical and numerically simpler than \textit{stat}DMFT,
does show this behavior opens the possibility of understanding it,
which is left as a direction of work to follow in the future.

This work was supported by CAPES (W.S.O.), CNPq and FAPEMIG (M.C.O.A.)
and NSF grant DMR-1005751 (V.D.).

\bibliographystyle{apsrev}
\bibliography{cttmt}

\end{document}